\documentclass[%
 reprint,
superscriptaddress,
 amsmath,amssymb,
 aps,
pra,
]{revtex4-2}

\usepackage{graphicx}
\usepackage{dcolumn}
\usepackage{bm}

\begin{document}

\preprint{APS/123-QED}

\title{Laser threshold magnetometry using green light absorption by diamond nitrogen vacancies in an external cavity laser}

\author{James L. Webb}
 \email{jaluwe@fysik.dtu.dk}
\affiliation{Center for Macroscopic Quantum States (bigQ), Department of Physics, Technical University of Denmark, Kgs. Lyngby, Denmark}%
\author{Andreas F. L. Poulsen}%
\affiliation{Center for Macroscopic Quantum States (bigQ), Department of Physics, Technical University of Denmark, Kgs. Lyngby, Denmark}%
\author{Robert Staacke}%
\affiliation{Division of Applied Quantum System, Felix Bloch Institute for Solid State Physics, \\ Leipzig University, 04103, Leipzig, Germany}%
\author{Jan Meijer}%
\affiliation{Division Applied Quantum System, Felix Bloch Institute for Solid State Physics, \\ Leipzig University, 04103, Leipzig, Germany}%
\author{Kirstine Berg-S{\o}rensen}
\affiliation{Department of Health Technology, Technical University of Denmark, Kgs. Lyngby, Denmark}%
\author{Ulrik Lund Andersen}
\affiliation{Center for Macroscopic Quantum States (bigQ), Department of Physics, Technical University of Denmark, Kgs. Lyngby, Denmark}%
\author{Alexander Huck}
\affiliation{Center for Macroscopic Quantum States (bigQ), Department of Physics, Technical University of Denmark, Kgs. Lyngby, Denmark}%

\date{\today}

\begin{abstract}
Nitrogen vacancy (NV) centers in diamond have attracted considerable recent interest for use in quantum sensing, promising increased sensitivity for applications ranging from geophysics to biomedicine. Conventional sensing schemes involve monitoring the change in red fluorescence from the NV center under green laser and microwave illumination. Due to the strong fluorescence background from emission in the NV triplet state and low relative contrast of any change in output, sensitivity is severely restricted by a high optical shot noise level. Here, we propose a means to avoid this issue, by using the change in green pump absorption through the diamond as part of a semiconductor external cavity laser run close to lasing threshold. We show theoretical sensitivity to magnetic field on the pT/$\sqrt{\rm Hz}$ level is possible using a diamond with an optimal density of NV centers. We discuss the physical requirements and limitations of the method, particularly the role of amplified spontaneous emission near threshold and explore realistic implementations using current technology. 
\end{abstract}
\maketitle

\section{Introduction}

Optical manipulation of material defects represents an ideal method for quantum sensing, exploiting properties such as entanglement and superposition \citep{Schleich2016}. The nitrogen-vacancy (NV) center in diamond, possessing long quantum coherence times at room temperature, has in particular drawn considerable interest \citep{Gruber1997,Doherty2013,Taylor2008}. Diamond is an ideal material for sensing, being mechanically hard, chemically stable, isotopically pure as well as biocompatible \citep{Schirhagl2014, Webb2020}. The negatively charged nitrogen-vacancy center (NV$^{-}$) has an energy level structure that results in optical properties that are highly sensitive to temperature \citep{Kucsko2013}, strain (pressure) \citep{Knauer2020}, electric field \citep{Dolde2011} and particularly magnetic field. Sensing is conventionally performed by detecting changes in the intensity of red fluorescence ($\approx$ 637-750 nm) under irradiation with green light and resonant microwaves via a process termed optically detected magnetic resonance (ODMR) spectroscopy \citep{Taylor2008, Hong2013,Rondin2014, Barry2020}. It can be done using a continuous wave (CW) method \citep{Webb2019}, or by using short laser and microwave pulses \citep{Wolf2015,Bucher2019}. 

However, measuring via red fluorescence suffers from two considerable physical limitations. First, the signal to be measured has a very low contrast on bright emission from decay in the NV$^{-}$ triplet state. Although for a single NV$^{-}$, spin dependent contrast can be up to 30$\%$ \citep{Osterkamp2019}, for a large ensemble of NVs suitable for a diamond sensor the contrast can be at most a few percent \citep{Levchenko2015,Wojciechowski2018}. Sensitivity is therefore limited by this low contrast and the high level of shot noise from the bright background rising from triplet state fluorescence emission. The second physical limitation is the high refractive index of diamond, which traps the majority of the fluorescence inside the diamond. Microfabrication schemes have been proposed to mitigate this issue, but have yet to deliver significant improvements \citep{Liu2019,Huang2019}. 

An alternative method is to use optical absorption of the pump light by the NVs. Previous work has used the change in green absorption in an optical cavity \citep{Ahmadi2018,Ahmadi2017} or by using changes in infrared (IR) absorption by the singlet state \citep{Jensen2014}. These schemes are technically demanding, requiring an optical cavity or unusual wavelength (1042 nm) laser. A promising alternative is laser threshold sensing \cite{Jeske2016}, using changes in optical absorption resulting from the parameter to be sensed (e.g. magnetic field or temperature) to push a medium across lasing threshold. This method eliminates the bright background that limits sensitivity using conventional fluorescence detection. A further attraction is wide applicability to any material with variable optical absorption, including a wider range of defects in diamond, SiC and 2D materials \citep{Jeong2019,Castelletto2020}. 

Building on the work by Dumeige et al.\ \cite{Dumeige2019} and our own previous work on diamond absorption magnetometry  \cite{Ahmadi2018,Ahmadi2017}, we outline here a scheme to use laser threshold sensing of magnetic field with green light in a standard external cavity laser. We show it is possible to achieve high sensitivity in the pT/$\sqrt{\rm Hz}$ range with realistic assumptions for key physical parameters. Our proposal differs from that of previous work by using simpler green pump absorption rather than IR absorption and by using an ordinary current driven laser diode/gain chip medium without the need for an additional pump laser. We show that this configuration, highly suitable for miniaturization, can deliver high sensitivity, and we discuss the key physics required to reach such sensitivity levels. Finally, we discuss and calculate limiting factors that may prevent these levels from being reached in practice. This includes factors that may not have been previously considered, such as amplified spontaneous emission near to lasing threshold.

\section{Methods}
\subsection{External Cavity Laser Model}

\begin{figure}
    \centering
    \includegraphics[width=7cm]{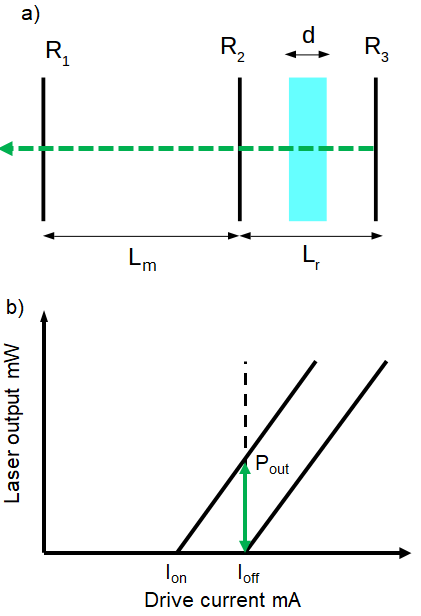}
    \caption{a) Schematic of the external cavity setup with a Fabry-Perot semiconductor laser diode of cavity length $L_m$ and end facet power reflectivities $R_1$ and $R_2$ coupled to an external cavity of length $L_r$ via mirror $R_3$ and diamond of thickness $d$. Laser emission (dashed line) is through $R_1$. We model our external cavity laser with the diamond as a single cavity with equivalent end reflectance $R_e$ and the diamond absorption loss $\alpha_d$ included in the total cavity loss $\alpha_t$, b) Simplified schematic of the laser threshold process, where a reduction in diamond absorption through application of resonant microwaves reduces the threshold current to $I_{th}^{\rm on}$, producing lasing output $P_{out}$ when driven at $I_{th}^{\rm off}$. }
    \label{fig:fig1A}
\end{figure}

We place the diamond into a standard external cavity laser setup as described schematically in Fig. \ref{fig:fig1A},a). This consists of a Fabry-Perot semiconductor laser diode or gain chip of length $L$ with end facet reflectivities $R_1$ and $R_2$ coupled to an external cavity formed by mirror $R_3$ via an external cavity of length $L_r$ containing the diamond of thickness $d$. We assume normal incidence and that transmission through the diamond is high, with minimal reflection from the diamond facets. For simplicity, we assume a single optical mode at a single wavelength.  We consider the optical loss due to absorption in a diamond of thickness $d$. The change in laser intensity $\hat{I}$ on a pass through the diamond is given by

\begin{equation}
\Delta \hat{I}=\hat{I_0}-\hat{I_0}e^{-\alpha_dz},
\end{equation}

where $\hat{I_0}$ is the intensity of the laser emission with no diamond present in the external cavity, $\alpha_d$ is the absorption coefficient in the diamond and $z$ is the path length taken within the diamond. For normal incidence $z$ = $d$ and the absorption coefficient can be derived from the rate equation model given in the following section or can be measured experimentally. For the semiconductor lasing medium between mirrors 1 and 2, we assume a total cavity loss $\alpha_t$, given by the sum of intrinsic cavity loss due to the gain medium $\alpha_c$ and losses from the mirrors and end facets $\alpha_m$, giving a total loss $\alpha_t$

\begin{equation}
\alpha_t = \alpha_c + \alpha_m = \alpha_c + \frac{1}{L_m}\ln\left (\frac{1}{\sqrt{R_1R_2}}  \right ).
\end{equation}

In order to simplify the analysis of the external cavity structure, we use the three mirror model  \cite{Cunyun2004, Petermann1988, deGroot1988,Li2017} to treat the complete diode/external cavity structure as a single cavity of length $L$ = $L_m$ + $L_r$, with mirror $R_2$ replaced by an effective reflectivity $R_e$, with the single cavity containing the optical losses of the external cavity and diamond, the internal losses of the gain medium in the laser diode and loss from the cavity through the mirrors. By assuming the losses due to the diamond are spread evenly throughout, we redefine the loss coefficient due to the diamond as $\alpha_{e} = (\alpha_{d}/L)d$, and our total cavity loss as

\begin{equation}
\label{eqn:at}
\alpha_t = \alpha_c + \alpha_e + \frac{1}{L}\ln\left (\frac{1}{\sqrt{R_1R_e}}  \right ),
\end{equation}

where $R_e=\left |r_e  \right |^{2}$ relates the power reflectivity to the complex field reflectivity, $r_e$. We use the model for the effective reflectivity by Voumard et al.  \cite{Voumard1977}, detailed further in the Supplementary Information. Neglecting phase components, at threshold $R_1 R_e e^{( \Gamma g-\alpha_t)2L}  = 1$, where $g$ = $g_{th}$ is the (threshold) gain coefficient. 

For the full structure, the rate equations for photon ($S$) and carrier ($N$) density are given by the standard equations for a laser diode as

\begin{equation}
\label{eqn:rate1}
\frac{dN}{dt}=\frac{I}{qV}-\frac{N}{\tau_N}-GS,
\end{equation}

and

\begin{equation}
\label{eqn:rate2}
\frac{dS}{dt}=GS-\frac{S}{\tau_P}+\frac{\beta N}{\tau_N}.
\end{equation}

Here $I$ is the drive current, $V$ the volume of the gain region, $G$ the gain of the lasing medium and $q$ the electronic charge. The term $GS$ arises from stimulated emission in the laser diode gain medium and $S/\tau_P$ includes the cavity loss from the mirrors, gain medium and diamond. Further, $\tau_P$ is the photon lifetime in the cavity and $\tau_N$ the carrier lifetime in the laser diode. Carriers are generated by a current $I$ in a volume $V$, where $V = L\times w \times t_h$, where $t_h$ is the thickness and $w$ the width of the laser diode active region. The term $\beta N/\tau_N$ relates to spontaneous emission, governed by the spontaneous emission factor $\beta$. 

We can define gain $G$ phenomenologically, in the form \cite{Allen1994}

\begin{equation}
\label{eqn:HS}
G = \Gamma g  = \Gamma a(N_{th}-N_{tr})(1-\epsilon S),
\end{equation}

where $\Gamma$ is the confinement factor and $\epsilon$ is the gain compression factor that phenomenologically accounts for effects such as spectral hole burning at higher optical power. The carrier density at transparency is given by $N_{tr}$. The rate equations for photon and carrier density can be solved for a steady state condition ($dS/dt$ = 0, $dN/dt$ = 0). For carrier density $N$ = $N_{th}$ close to $N_{tr}$ and neglecting spontaneous emission ($\beta$ = 0), the gain balances the cavity loss. The factor $a$ is the differential gain coefficient, a material specific property defining how well the semiconductor can generate carriers for population inversion. Equation (\ref{eqn:HS}) is valid for heterostructure laser diodes and certain quantum well structures where the threshold is close to the transparency carrier density. Unless otherwise stated, we use the Eq. (\ref{eqn:HS}) model in this work.

Using Eqs. (\ref{eqn:rate1}-\ref{eqn:HS})  at lasing threshold, where $S = 0$, $G$ = 1/$\tau_p$ and $\Gamma g_{th}$ = $\alpha_t$ we can derive an equation for carrier density at threshold $N_{th}$

\begin{equation}
\label{eqn:nth}
N_{th}=N_{tr}+\frac{\alpha_t}{\Gamma a},
\end{equation}

and inserting this result into the rate equation for carrier density - Eq. (\ref{eqn:rate1}) - allows us to calculate the threshold current

\begin{equation}
\label{eqn:Ith}
I_{th}=\frac{qV}{\eta_i \tau_{N}}N_{th} = \frac{qV}{\eta_i \tau_{N}}\left ( N_{tr} + \frac{\alpha_t}{\Gamma a} \right ).
\end{equation}

Here we introduce the quantum efficiency of carrier to photon conversion $\eta_i$. By using Eq. (\ref{eqn:Ith}) in the rate equations at $I>I_{th}$, we can calculate the photon density at any current above the lasing threshold. We can then calculate the laser light power that can be emitted from the left hand side mirror $R_1$ using the factor $\eta_o$, the output coupling efficiency, which is defined as the ratio of photons lost through the mirror $R_1$ to the total cavity loss $\alpha_t$ = $\alpha_m + \alpha_c + \alpha_e$

\begin{equation}
P_{out}=\eta_o\frac{hc}{\lambda \tau_p}\frac{V}{\Gamma}S,
\end{equation}

where $V/\Gamma$ is the effective mode volume of the cavity, $\lambda$ the wavelength and $h$ and $c$ Plank's constant and the speed of light respectively. In the limit of $\epsilon S \rightarrow 0$ where there is no limiting effect on the gain, the power output can be rewritten directly in terms of the threshold current

\begin{equation}
\label{eqn:pow1}
P_{out}=\eta_o\eta_i\frac{hc}{q\lambda}(I-I_{th}).
\end{equation}

In both of these expressions
\begin{equation}
\eta_o=\frac{\alpha_{m_1}}{\alpha_{m}+\alpha_{c}+\alpha_{e}}=\frac{ln\frac{1}{\sqrt{R_1}}}{ln\frac{1}{\sqrt{R_1R_e}}+(\alpha_e+\alpha_c)L}.
\end{equation}

For larger finite values of $\epsilon$ well above threshold or including finite spontaneous emission through nonzero $\beta$, we can numerically solve the steady state rate equations (Eqs. (\ref{eqn:rate1}) and (\ref{eqn:rate2})) to calculate $N$, $S$ and the laser power output. 

The total cavity absorption $\alpha_t$ will change when microwaves are applied to the diamond at frequency equal to the splitting of the NV triplet ground state levels, reducing the lasing threshold current $\Delta I_{th}$ = $I_{th}^{\rm off}$ - $I_{th}^{\rm on}$, where $I_{th}^{\rm on}$ is the threshold current \textit{on} microwave resonance, and $I_{th}^{\rm off}$ the threshold current \textit{off} resonance. By running at drive current equal to $I_{th}^{\rm off}$, laser output is generated only while on microwave resonance. This is shown schematically in Fig. \ref{fig:fig1A},b).

\subsection{Absorption model}

We use the rate equation model from  \cite{Robledo2011} in order to calculate the optical absorption of green pump light by the diamond and the maximum change in absorption when on microwave resonance. The parameters we use for the transition rates are the same as those in  \cite{Dumeige2019}, derived from  \cite{Tetienne2012, Acosta2010, Meirzada2018, Wee2007}. We calculate the normalized occupancies of each energy level with microwaves supplied $n_i^{\rm on}$ and without microwaves $n_i^{\rm off}$, where $\sum_i n_i$ = 1 and index $i$ = 1-8, where i=1 refers to the $m_s$=0 ground state level, i=2 the $m_s$=$\pm$1 ground state levels, i=3,4 the spin triplet excited states, i=5,6 the spin singlet shelving states and i=7,8 the ground and excited state of the NV$^0$. We define a total NV$^{-}$ density $N_{NV}$ in ppm. Off resonance, the total number density of NV$^{-}$ in each state $N^{\rm off}_i$ are given by

\begin{equation}
N^{\rm off}_i = N_{NV}\frac{n_i^{\rm off}}{\sum_i n_i^{\rm off}}.
\end{equation}

We define a measurement axis along one of the 4 possible crystallographic axes for the NV. We calculate that when microwaves are applied, we drive only the NVs aligned along one axis such that the total number density on resonance $N^{\rm on}_i$ is given by 

\begin{equation}
N^{\rm on}_i = \frac{1}{4}N_{NV}\frac{n_i^{\rm on}}{\sum_i n_i^{\rm on}} + \frac{3}{4}N_{NV}\frac{n_i^{\rm off}}{\sum_i n_i^{\rm off}}.
\end{equation}

We calculate the change in intensity on a single pass when on and off microwave resonance as

\begin{equation}
\begin{split}
\hat{I_{\rm on}}=\hat{I_0}e^{-\alpha_{\rm on} d}, \\
\hat{I_{\rm off}}=\hat{I_0}e^{-\alpha_{\rm off} d},
\end{split}
\end{equation}

where $d$ is the thickness of the diamond and the absorption coefficient $\alpha$ on and off resonance is given by

\begin{equation}
\alpha^{\rm on} =\sigma_g(N^{\rm on}_1 + N^{\rm on}_2)+\sigma_{g0}N^{\rm on}_7+\sigma_e(N^{\rm on}_3+N^{on}_4)+\sigma_rN^{\rm on}_8, 
\end{equation}

\begin{equation}
\begin{split}
\alpha^{\rm off} =\sigma_g(N^{\rm off}_1 + N^{\rm off}_2)+\sigma_{g0}N^{\rm off}_7+\sigma_e(N^{\rm off}_3+ \\
 N^{\rm off}_4)+\sigma_rN^{\rm off}_8.
\end{split}
\end{equation}

Here $\sigma_g$ and $\sigma_{g0}$ are respectively the absorption cross sections of green light for NV$^{-}$ and NV$^0$ and $\sigma_e$, and $\sigma_r$ the ionisation cross sections for transfer between the charged and uncharged defect state. This allows us to calculate the change in absorption when the diamond is present without microwaves $\hat{I_{\rm off}}/\hat{I_0}$, the change when driven on microwave resonance $\hat{I_{\rm on}}/\hat{I_0}$ and the change between these, which we term the absorption contrast 

\begin{equation}
C = (\hat{I_{\rm off}}/\hat{I_0}) - (\hat{I_{\rm on}}/\hat{I_0}). 
\end{equation}

\subsection{Key Physical Parameters}

The key physical parameters of the model can be divided into those that are intrinsic to the semiconductor gain medium, those intrinsic to the diamond and those defined by the setup. Examples of the latter include the mirror reflectivities $R_1$, $R_2$, $R_3$, the cavity length $L$ and any other losses, such as reflection out of the cavity or from absorption by other optical components such as lenses, included in the cavity loss factor $\alpha_{c}$. These factors will also influence the photon lifetime in the cavity $\tau_{P}$. The maximum Rabi frequency  $\Omega_R$ that can be reached also depends on microwave power and how well the microwaves can be coupled into the diamond.

\begin{figure}
\includegraphics[width=8.6cm]{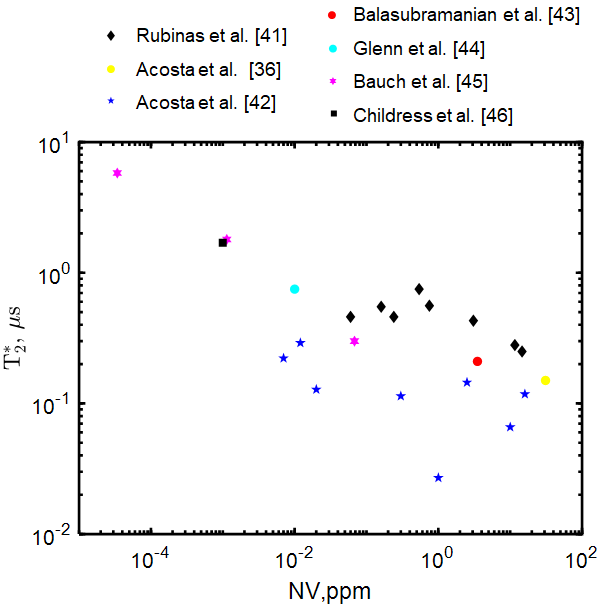}
\caption{Dephasing time $T_2^{*}$ vs NV$^{-}$ density $N_{NV}$ where both values are given in other works (citations in the main text). $T_2^{*}$ in those with low NV concentration are limited by interaction with ${}^{13}$C spin, with the highest values given by diamonds isotropically purified with ${}^{12}$C spin during growth. $T_2^{*}$ in those with high NV concentration is limited by dipolar interaction between defects, including other subsitutional nitrogen defects such as P1 centers.  (Note: the NV$^{-}$ density for the work by Childress et al. is an upper estimate made here assuming a 10$\%$ NV$^{-}$ fraction; total substitutional nitrogen content for this diamond was given as $\leq$0.1 ppm.}
    \label{fig:t2starfig}
\end{figure}

The parameters which are intrinsic to the diamond are the diamond thickness $d$, $NV^{-}$ density $N_{NV}$, ensemble dephasing time $T_2^{*}$ defining the ODMR linewidth and absorption contrast $C$ arising from changes in pump absorption on or off microwave resonance. These factors define the diamond absorption factor $\alpha_d$.  

A number of these parameters are interrelated. The ODMR linewidth is proportional to the inverse of $T_2^{*}$, which in turn is dependent on $N_{NV}$ concentration in the limit of high nitrogen content and the abundance of ${}^{13}$C for low nitrogen content \cite{Jahnke2012}. There is also a dependence on other material properties such as strain \cite{Kehayias2019}, which makes the relationship between the parameters difficult to determine. We therefore consider values in the experimental literature as a guide. Fig. \ref{fig:t2starfig} shows a plot of $T_2^{*}$ versus NV$^{-}$ density $N_{NV}$ for a range of diamonds from the literature  \cite{Rubinas2018,Acosta2010,Acosta2009,Balasubramanian2019, Glenn2018, Bauch2018, Childress2006}. Typical NV$^{-}$ densities range from 0.1ppb up to tens of ppm\citep{Su2013}. In general, $T_2^{*}$ $<$ 1 $\mu$s for samples with natural (1.1$\%$) ${}^{13}$C content \citep{bauch2019decoherence,Maze2009}. Experiments typically realize Rabi frequencies $\Omega_R$ of 1-5 MHz, with up to 10 MHz using optimal antenna geometries  \cite{Yaroshenko2020}. 

\begin{table}
\begin{tabular}{ |c|c|c| } 
 \hline
 Parameter & Range & Ref.\\ 
\hline 
 Transp. carrier density, $N_{tr}$ & $3\times10^{18}$-$2\times10^{19}$cm$^{-3}$ &  \cite{Fu2019, Nakamura1999, thesis6083,TabatabaVakili2020,Farrell2011} \\ 
 Carrier lifetime, $\tau_{N}$  & 1-5ns &  \cite{Lutgen2010, Nakamura1997}\\ 
 Differential gain factor, $a$ & 10$^{-17}$ - 10$^{-22}$m$^2$ &  \cite{AlGhamdi2019,Frost2013} \\ 
 Confinement factor, $\Gamma$ & 0.01-0.1 &  \cite{Staczyk2013,Zhang2009} \\ 
 Spont. emission factor, $\beta$ & 10$^{-5}$-10$^{-2}$ &  \cite{Cassidy1991,Scheibenzuber2011} \\ 
 \hline
  \end{tabular}
	\caption{Typical ranges for the key semiconductor gain medium parameters. Here $N_{tr}$, $\tau_{N}$ and $a$ are taken for typical III-nitride semiconductors. The range for $\Gamma$ is given for laser diodes with a thin (sub-$\mu$m) active layer and is typically no more than a few percent. The range of $\beta$ is given for literature values for a range of laser diodes where confinement is not deliberately sought e.g. microcavities, where values several orders of magnitude higher than the given range are possible \citep{Kreinberg2017}.}
  \label{Tab:tablediode}
\end{table}

Those parameters intrinsic to the laser diode/gain chip used are the carrier density at transparency $N_{tr}$, the gain compression factor $\epsilon$ that arises from effects that limit the gain well above threshold, the differential gain coefficient $a$ that relates gain and carrier density, the threshold carrier lifetime $\tau_{N}$ , the confinement factor $\Gamma$, the volume of the gain medium $V$ and the spontaneous emission factor $\beta$. For our gain medium we take a III-V semiconductor heterostructure device, such as the nitride compounds capable of emission at green wavelengths (e.g. InGaN) \cite{Yang2017}. Table \ref{Tab:tablediode} shows a typical range of values for each of these parameters. We take the typical ranges shown based on experimental results from different structures (quantum well, vertical cavity) and from calculations based on bulk material properties such as effective mass. $N_{tr}$ effectively defines the size of the lasing threshold current $I_{th}$. The desired change in threshold current on change in absorption factor $\alpha_{t}$ is defined in particular by $\Gamma$ and the gain coefficient $a$ in Eqs. (\ref{eqn:at}) and (\ref{eqn:nth}).

\section{Results}

\subsection{Absorption contrast}

\begin{figure*}
    \includegraphics[width=17.2cm]{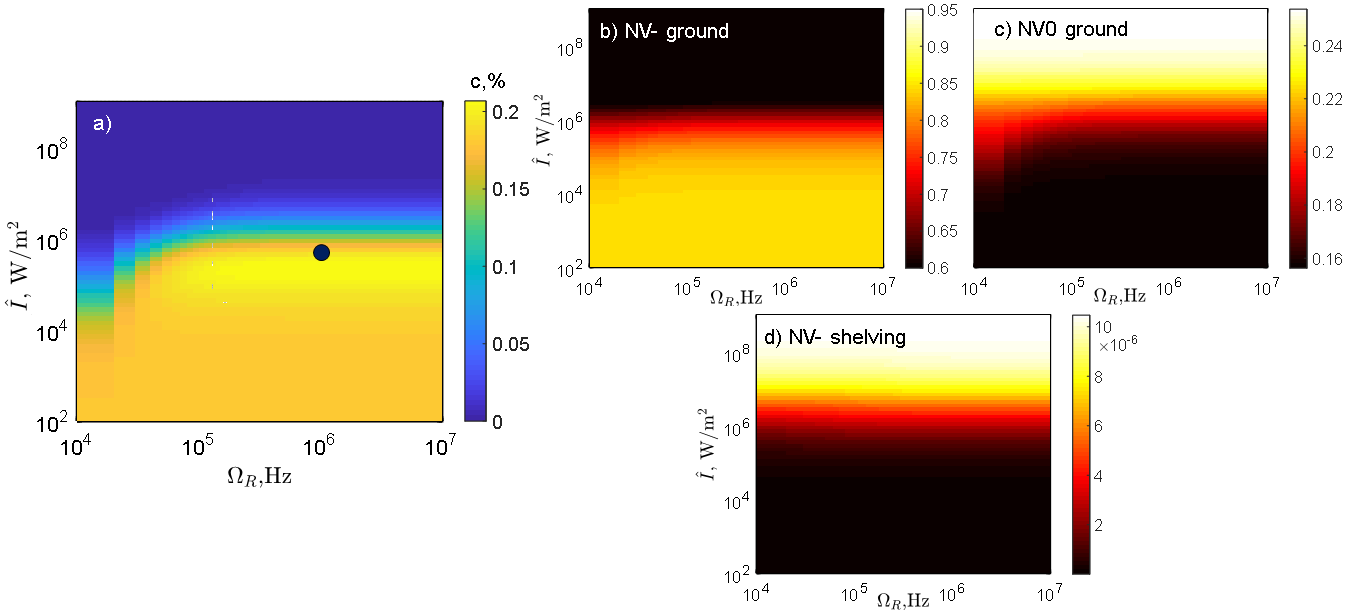}
    \caption{a) Absorption contrast percentage calculated from the rate model for Diamond D3. This is the maximum change in absorption between on microwave resonance and off microwave resonance as a function of Rabi frequency $\Omega_R$ and laser intensity $\hat{I}$ in W/m$^2$. b)-d) Normalized level occupancy for the NV$^{-}$ triplet ground state, responsible for green absorption, the uncharged NV$^{0}$ defect state and the NV$^{-}$ singlet state. At high laser intensities, population transfer to NV$^{0}$ limits the achievable absorption contrast. For reference, 10 mW of laser power with a 1 mm diameter circular beam on the diamond gives an intensity $\hat{I}$ = $10^4$ W/m$^2$. The black spot indicates the Rabi frequency and power for calculations later in this work.}
    \label{fig:contrastfigA} 
\end{figure*}

We first calculate from the rate model the fraction of incident pump light which is absorbed by the diamond and the change in this absorption ($C$) when on microwave resonance. We choose to model three different diamonds covering different regimes: D1, D2 and D3 with parameters (NV density and $T_2^{*}$) representative of the values seen in the literature (Fig. \ref{fig:t2starfig}). For Diamond D1 we choose a low NV$^{-}$ concentration $N_{NV}$ = 0.001 ppm, high $T_2^{*}$ = 5 $\mu$s, representative of ${}^{12}$C enriched diamonds. For Diamond D2 we choose a medium NV$^{-}$ concentration $N_{NV}$ = 0.1 ppm, $T_2^{*}$ = 0.75 $\mu$s, representative of CVD-grown diamond with natural ${}^{13}$C abundance. For Diamond D3 we choose $N_{NV}$ = 10 ppm, $T_2^{*}$ = 0.1 $\mu$s, characteristic of high nitrogen content high-pressure high-temperature (HPHT) diamond. We use a diamond thickness $d$ = 500 $\mu$m for all, representative of commercially available single crystal plates. 

Using the rate model, we can calculate the absorption of light incident on the diamond 1 - $\hat{I_{\rm off}}$/$\hat{I_0}$ where $\hat{I_0}$ is intensity of the incident light and $\hat{I_{\rm off}}$ the intensity of the light after the diamond (without supplying microwaves). The calculated absorption for Diamonds D1 - D3 is 0.015$\%$, 1.498$\%$ and 77$\%$, as expected from increasing NV density. We can also calculate the change in absorption when on and off microwave resonance. This is shown in Fig. \ref{fig:contrastfigA} as absorption contrast $C$ for D3 as a function of microwave drive power (as Rabi frequency $\Omega_R$) and laser output (as intensity). The equivalent plots for D1 and D2 are given in the Supplementary Information. Maximum $C$ = 0.22$\%$ for D3 and lowest for D1 with the lowest NV density with $C$ = 10$^{-4}\%$. This contrast is comparable to our previous absorption experiments using a diamond with equivalent ppb-level NV$^{-}$ density \cite{Ahmadi2018}. We note that at Rabi frequencies above 100 kHz and laser outputs  above 10$^6$ W/m$^2$ the absorption contrast begins to drop. This results from depopulation of the triplet ground state ${}^{3}$A$_2$ (normalized occupancy in Fig. \ref{fig:contrastfigA},b) in favor of the NV$^{0}$ (Fig. \ref{fig:contrastfigA},c) and the singlet shelving state (Fig. \ref{fig:contrastfigA},d). However, since we aim to operate near the lasing threshold, laser intensity will be low in our scheme, avoiding this issue and ensuring we remain in the region of highest contrast. 

\begin{figure}
    \centering
    \includegraphics[width=8.6cm]{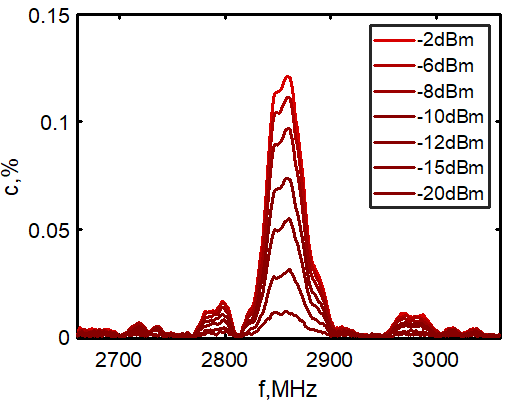}
    \caption{Experimental absorption contrast percentage as a function of microwave drive frequency and microwave power before the amplifier (Minicircuits ZHL-16W), measuring through the diamond with 100 mW of laser light ($\hat{I}$ = 2.5x10$^{4}$ W/m$^{2}$). The sample was used to test the absorption model using estimates of NV density and $T_2^{*}$ from the observed linewidth (values in the main text). Note: due to input loss we remain below the maximum gain threshold of the amplifier for all microwave powers shown, which is exceeded at +3dBm. }
    \label{fig:leipzig}
\end{figure}
 
To further validate the absorption modeling, we have also measured diamond absorption on a high density sample consisting of a 1 mm  thick HPHT diamond with 200 ppm nitrogen content, irradiated with 10 MeV electrons and annealed at 900 $^{\circ}$C. The estimated NV content for this sample was 10-20 ppm. The absorption contrast for this sample is shown in Fig. \ref{fig:leipzig}. The sample was found to be moderately polycrystalline and was therefore measured without an offset field to produce a single central dip in fluorescence, with a number of satellite features resulting from the polycrystalinity and residual magnetic field in the laboratory. Here total off resonance diamond absorption was 90$\%$ of incident pump light and maximum absorption contrast $C$ = 0.13$\%$. For comparison to experiment, we model absorption contrast with our model with NV density of 15 ppm, $T_2^{*}$ = 100 ns, derived from an estimate of the resonance linewidth, an estimated Rabi frequency of 1 MHz and the same 100 mW laser power as used experimentally. This gives a total off resonance absorption of 89$\%$ of the pump light and absorption contrast of $C$ = 0.14$\%$, in good agreement with our measurements. 

\subsection{Change in Threshold Current}

\begin{figure}
    \centering
    \includegraphics[width=8.6cm]{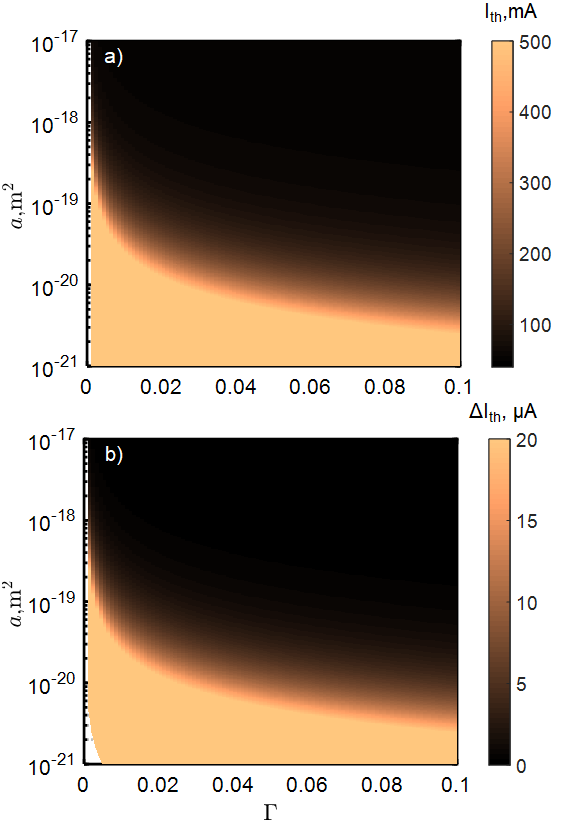}
    \caption{a) Threshold current as a function of differential gain factor $a$ and confinement factor $\Gamma$ for Diamond D3. b) Change in threshold current due to the diamond absorption contrast $C$ = 0.02$\%$ for Diamond D3. Here external cavity length $L_r$ was 10 mm and output mirror reflectivity $R_1$ = 0.9.}
    \label{fig:thresholdfigA}
\end{figure}

We first calculate the lasing threshold current $I_{th}$ with the diamond absent from the cavity. To do this we fixed some of the parameters of the semiconductor gain medium. We choose a transparency carrier density of $N_{tr}$ = 1$\times$10$^{25}$ m$^{-3}$ in the range typical for InGaN laser structures \cite{Fu2019}, a gain region volume of $V$ = 1.25$\times$10$^{-16}$ m$^3$ (25 $\mu$m x 100 nm x 100 $\mu$m). For zero total cavity absorption $\alpha_t$ = 0, a typical differential gain factor $a$ = 5$\times$10$^{-20}$ m$^2$, a confinement factor $\Gamma$ of 2$\%$ and a carrier lifetime $\tau_N$ = 4 ns, giving a reasonable lasing threshold current of 50 mA \cite{Yang2017}. We take the relation between the gain and the carrier density to be linear, with the carrier density close to transparency. We make the simplifying assumption that due to the low power, running close to lasing threshold we do not encounter gain compression effects, such that the factor $\epsilon \rightarrow$ 0. We also initially make the simplifying assumption that the spontaneous emission rate is low, with $\beta\rightarrow$0 (the importance of this second assumption will be tested in the final section of this work). These assumptions allow the threshold current $I_{th}$ to be calculated easily from Eq. (\ref{eqn:Ith}). We define $L_r$ = 10 mm, sufficient to include the diamond and any necessary optics in a practical implementation. We set mirror reflectivity $R_3$ = 0.99 and collect laser output from transmission through mirror $R_1$. We calculate reflectivity $R_2$ from the Fresnel equations assuming an In$_{x}$Ga$_{1-x}$N/air interface with refractive index $n$ $\approx$ 2.6-2.9 for In$_{x}$Ga$_{1-x}$N \citep{Anani2007}.

We impose two feasibility limits on the threshold current $I_{th}$. The first is that it should not exceed 300 mA, based on the limits discussed in technical documentation, in order to maintain thermal stability and for practical heatsinking for a miniaturized diode/gain chip medium. The second is that the change in the threshold caused by the diamond absorption must exceed the shot noise of the drive current. From Fig. \ref{fig:thresholdfigA}, we can see that these are mutually exclusive objectives. Setting an absorption contrast $C$ = 0.2$\%$ (Diamond D3) and output mirror reflectivity $R_1$ = 0.9 in order to achieve laser output while keeping threshold current reasonably low, a low confinement factor $\Gamma$ and high differential gain coefficient $a$ result in the highest change in threshold current $\Delta I_{th}$ and thus strongest effect for sensing, but for very high $I_{th}$. Conversely, a higher value of $\Gamma$ or lower $a$ gives lower $I_{th}$, but $\Delta I_{th}$ shifts which are too small to be resolved. 

Although N$_{tr}$ is a factor usually defined by the semiconductor material, we note that the other parameters here which define $I_{th}$, $\Gamma$, $a$ and total cavity loss $\alpha_t$ including the mirror reflectivity and cavity output through $R_1$, are all factors which are well understood and can be controlled and optimized at either the semiconductor growth stage or in the external cavity design. 

\subsection{Simulated ODMR}

\begin{figure}
    \centering
    \includegraphics[width=8.6cm]{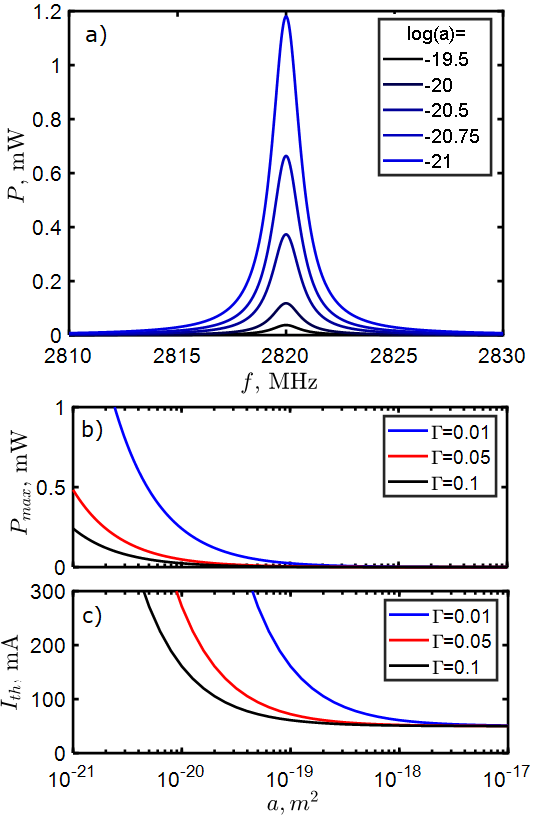}
    \caption{a) Simulated ODMR for Diamond D3 at a range of differential gain factors $a$=10$^{-21}$ $\rightarrow$ 10$^{-19.5}$ m$^{2}$ (exponents given in legend) measured by calculating external cavity laser output power $P$ as a function of microwave frequency for a Lorentzian lineshape transition centered at 2.83 GHz and of linewidth defined by f$_l$ = $\frac{1}{\pi T_2^{*}}$ = 3.2 MHz. b) Maximum laser power output $P_{max}$ on resonance as a function of gain coefficient $a$ for confinement factor $\Gamma$ = 0.01, 0.05, 0.1, c) the lasing threshold current for $I_{th}$ $<$ 300 mA, for the same three values of $\Gamma$. }
    \label{fig:simODMR}
\end{figure}

Here we calculate the ODMR spectrum that would be produced from the external cavity laser. We model a single microwave resonance from a single $m_s$ = 0 $\rightarrow$ $m_s$ = $\pm$1 transition using a Lorentzian lineshape typical of ODMR for diamond \cite{Levchenko2015}. We center our resonance at 2.82 GHz, replicating an ODMR resonance feature associated with a single NV axis, split from resonance features from other axes by an arbitrary weak DC offset magnetic field. The maximum amplitude is defined by the maximum change in threshold current between on and off microwave resonance and full width half maximum linewidth f$_l$. For simplicity we assume that we can reach the pulsed readout linewidth defined by $T_2^{*}$.  We calculate the external cavity laser output power using Eq. (\ref{eqn:pow1}). Fig. \ref{fig:simODMR},a) shows the simulated ODMR for Diamond D3, with a laser power output in the mW range for reasonable values of $\Gamma$ $<$ 0.1 and $a$ = 10$^{-17}$-10$^{-21}$ m$^2$. The equivalent plots for Diamond D1 and D2 are given in the Supplementary Information, with maximum power outputs in the range of nW and $\mu$W respectively. Unlike for conventional red fluorescence ODMR, the spectrum using this method is a peak at microwave resonance with zero background, rather than a small percentage change on a bright background.
\subsection{Magnetic Field Sensitivity}

\begin{figure}[h]
    \centering
    \includegraphics[width=8.6cm]{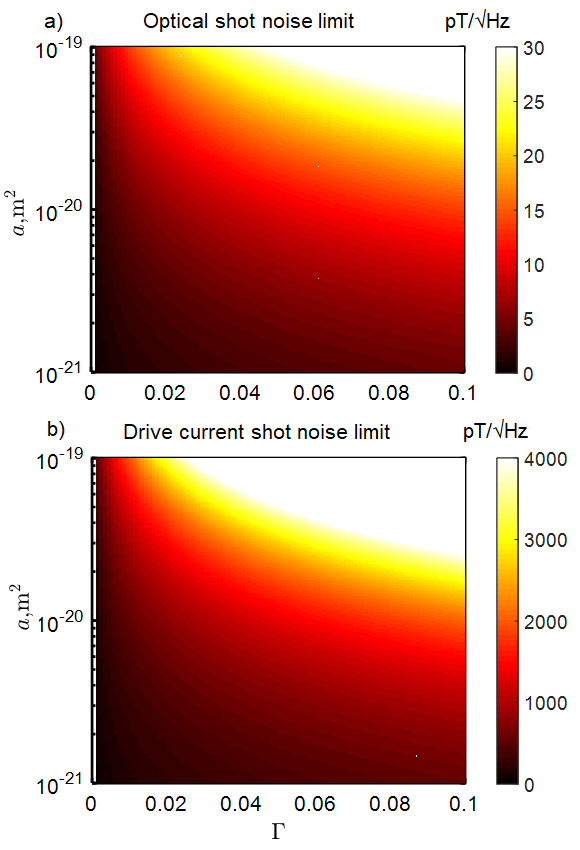}
    \caption{a) Optical shot noise limited sensitivity for Diamond D3 within a viable range for diode parameters $a$ and $\Gamma$. Sensitivity is in the picotesla range, enabled by the elimination of the high noise from the background in the conventional fluorescence detection scheme. b) The ultimate sensitivity limit imposed by shot noise on the laser drive current, worse by up to 2 orders of magnitude. These plots give no consideration for practical viability, with threshold currents $>$4 A at low $\Gamma$.}
    \label{fig:fieldsens}
\end{figure}

We calculate sensitivity to magnetic field by taking the background noise level, dividing by the maximum ODMR slope and by assuming a maximum frequency shift of 28 Hz $\approx$ 1 nT  \cite{Webb2019}. In our model, the primary sources of noise are readout from the photodetector and the noise on the drive current. The ultimate limit on both of these is shot noise of the output laser light and the drive current shot noise. We make no account for other direct sources of noise which are difficult to quantify, such as vibration or temperature fluctuations.  Fig. \ref{fig:fieldsens} shows a plot of sensitivity for Diamond D3 versus laser diode parameters for a) the optical shot noise and b) drive current shot noise limited regimes, with best sensitivity of 50 pT/$\sqrt{\rm Hz}$ when limited by the shot noise of the drive current. We note that in practice the shot noise limited operation may be experimentally difficult to realize and include an estimate based on a commercial current source with ppm-level noise in the Supplementary Information. 

\begin{figure}[h]
    \centering
    \includegraphics[width=8.6cm]{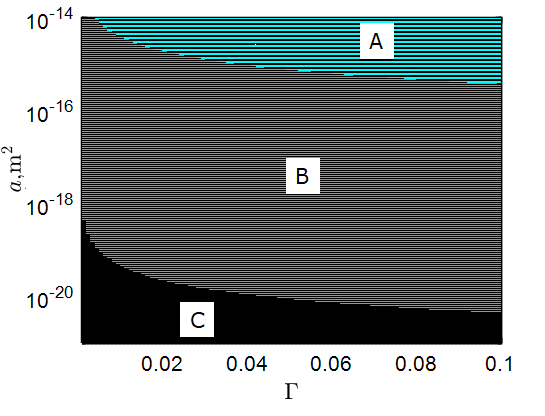}
    \caption{Regions where the sensor can and cannot operate due to imposed limitations. In Regions A and C, operation is constrained by having a change in threshold less than the shot noise of the drive current and $I_{th}$ $>$ 300 mA respectively. In Region B, operation is possible.}
    \label{fig:banned}
\end{figure}

The noise limitations as a function of diode parameters are highlighted in Fig. \ref{fig:banned} for Diamond D3. Here the regions A and C represent where operation is noise limited and the region B represents the region in which the system can operate. In region A, the change in threshold current is less than the shot noise of the laser drive current ($\Delta I_{th}$ $<$ $I_{sh}$). In region C, the threshold current $I_{th}$ $>$ 300 mA exceeds a reasonable maximum drive current in order to maintain thermal stability. The limitations we impose mean Diamond D1 or D2 have no viable operating region. For completeness, their sensitivity plots are included in the Supplementary Information. 

\begin{figure}[h]
    \centering
    \includegraphics[width=8.6cm]{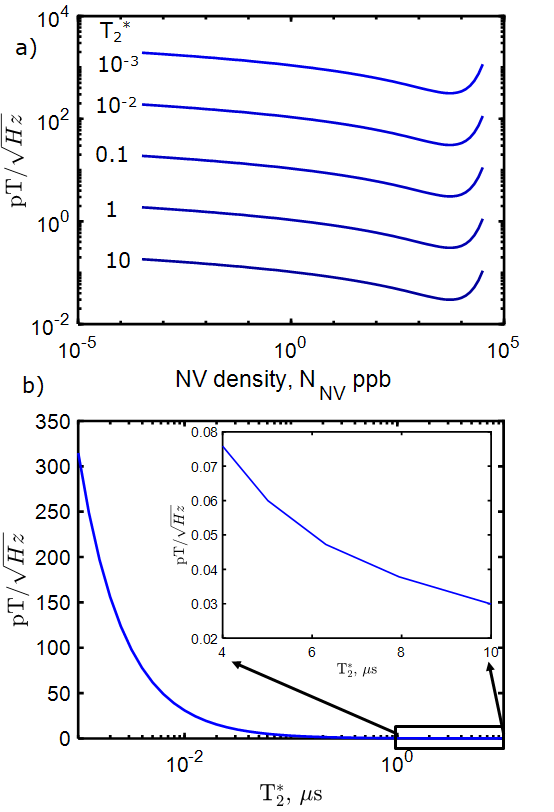}
    \caption{Best field sensitivity optimizing variables listed in the main text as a function of a) NV$^{-}$ density and b) $T_2^{*}$ in $\mu$s, with the inset showing a zoomed plot at the highest simulated values of $T_2^{*}$. The best sensitivity was observed at the highest $T_2^{*}$, for NV$^{-}$ density 10$^{4}$ ppb. Above this the total absorption for the diamond was too high, limiting laser output and sensitivity. }
    \label{fig:fullopt}
\end{figure}

By solving the rate model and calculating for laser diode output, we can calculate sensitivity to magnetic field for any valid physical parameters of the system, regardless of whether a diamond can be created with the requisite properties. This includes whether a value of $T_2^{*}$ can be realized for a corresponding $N_{NV}$, making no assumption regarding the relation between these parameters, or whether $N_{NV}$ can be realized experimentally. Here we choose parameters $R_1$, $\Gamma$, $T_2^{*}$, $a$ and NV$^{-}$ density $N_{NV}$ as optimization variables, while fixing diamond thickness ($d$ = 500 $\mu$m), Rabi frequency ($\Omega_R$=1 MHz), laser beam width (0.5 mm), power (200 mW) and mirror reflectivities. We limit our laser power to 200 mW based on our rate model calculations, to ensure the majority of light is absorbed by the NV$^{-}$ defects. We optimize using standard gradient descent methods. Fig. \ref{fig:fullopt} shows a plot of optical shot noise limited field sensitivity as a function of $T_{2}^{*}$ and NV density. Sensitivity increased with higher $T_2^{*}$ as would be expected, with maximum sensitivity at NV density of 10$^{4}$ ppb, above which high overall absorption by the diamond acted to excessively reduce laser output. Sub-picotesla level sensitivity is predicted for $T_2^{*}$ $>$ 1 $\mu$s (0.3-0.02 pT/$\sqrt{\rm Hz}$ for $T_2^{*}$ = 1-10 $\mu$s). Here the optimal parameters were $a$ = 1.6x10$^{-20}$ m$^2$, $R_1$ = 0.154 and $\Gamma$ = 0.025. These are parameters within the achievable range for a semiconductor gain medium (see Table \ref{Tab:tablediode}). 

\begin{figure}[h]
    \centering
    \includegraphics[width=8.6cm]{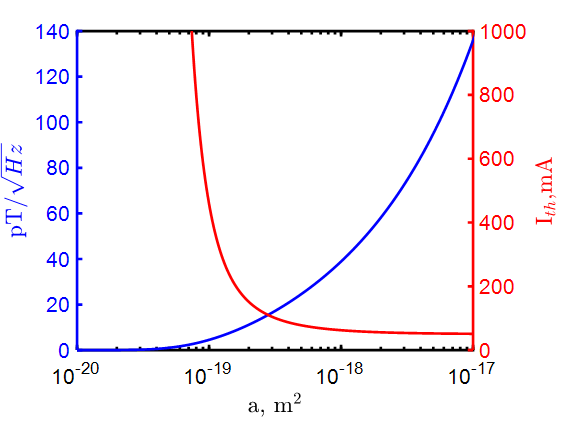}
    \caption{Calculated magnetic field sensitivity and laser diode threshold current as a function of differential gain factor $a$ using an empirical model for a quantum well laser diode. The lasing threshold current increases such that sub-picotesla sensitivity is not reached at a feasible threshold current ($<$300 mA). Here we take confinement factor $\Gamma$=0.01 as an example of the low values typical of a quantum well laser diode.}
    \label{fig:QW}
\end{figure}

We note that in general, the highest sensitivity is realized for the lowest differential gain factor $a$. A standard laser diode demands a large $a$, maximizing gain vs carrier density (steeper output power vs drive current slope). Our scheme requires the reverse: that a small change in gain produced by the diamond on/off microwave resonance results in a large change in $N_{th}$ and $I_{th}$. In this respect, a quantum well structure with a flatter logarithmic relation between gain and carrier density would seem preferable. However, as we demonstrate in Fig. \ref{fig:QW}, using our model, with modifications to the phenomenological description of the medium gain (detailed in the Supplementary Information) and with the same optimization methodology as above results in the threshold current exponentially exceeding drive current feasibility limits before sub-picotesla/$\sqrt{\rm Hz}$ sensitivity is reached, for any typical value for $\Gamma$ in the low percentage range. 

\subsection{Effect of spontaneous emission}

\begin{figure}[h]
    \centering
    \includegraphics[width=8.6cm]{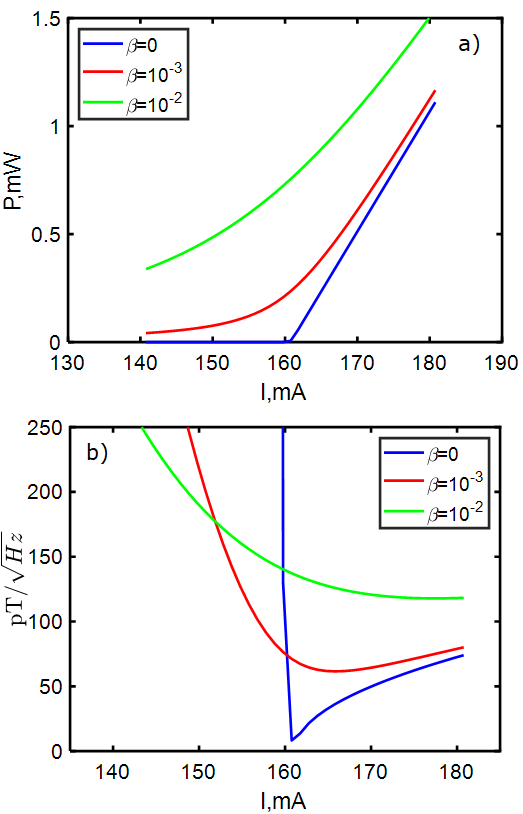}
    \caption{a) External cavity laser output $P$ as a function of semiconductor laser medium drive current $I$, varying spontaneous emission factor $\beta$. The result of increasing $\beta$ is that there is no longer a sharp lasing cut-on at threshold. b) shows the effect on the achievable sensitivity of this effect, with sensitivity considerably reduced by up to 2 orders of magnitude for $\beta$ = 10$^{-2}$.}
    \label{fig:spont}
\end{figure}

In the previous sections and past literature, the physical role of spontaneous emission in the semiconductor gain medium used was not considered. In order to maximize sensitivity, it is necessary to operate at or close to the off-resonance lasing threshold. Without spontaneous emission, this can be treated as a step cut-on, with zero or near-zero emission before lasing begins at $I_{th}$. With spontaneous emission included, modeled by finite $\beta$ in Eqs. (\ref{eqn:rate1}) and (\ref{eqn:rate2}) above, the power-current relationship close to threshold instead follows a shallow curve, resulting from weak amplification of spontaneous emission near threshold producing light emission below $I_{th}$. Fig. \ref{fig:spont},a) shows this effect for varying $\beta$. This acts to severely limit sensitivity (Fig. \ref{fig:spont},b) by reducing the contrast and adding background shot noise. Typical values of $\beta$ range from 10$^{-3}$ to 10$^{-5}$, depending on laser diode structure. We estimate approximately an order of magnitude worse sensitivity at the low end of this range than with $\beta$ = 0. 

\section{Conclusion}

In this work, we propose a scheme for laser threshold sensing using an external cavity laser configuration with a current driven semiconductor lasing medium. Using the change in lasing threshold, light emission only occurs on microwave resonance. This eliminates the bright background that limits sensitivity using conventional red fluorescence emission. Predicted sensitivities for magnetometry with realistic cavity parameters and intrinsic material parameters are in the pT/$\sqrt{\rm Hz}$ range, offering a route to improvement over existing methods. Our model has limitations: we base our calculations on emission into a single laser mode and do not calculate the dynamics of the system, such as rapid switching in a pulsed operation scheme. Although beyond the focus of this work, we note that the latter may be a promising route for future investigation. A scheme where the laser medium could be initially pumped and then the pump shut off while retaining population inversion during sensing, typical of a Q-switched setup, would only be limited by the optical shot noise of any emitted laser light. This is however challenging to achieve for a semiconductor laser due to the short excited state (carrier) lifetime. 

A key physical limitation of any laser threshold scheme is the role of amplified spontaneous emission. This blurs the sharp lasing transition, giving nonzero light emission even below threshold and compromising sensitivity. A broad transition can be avoided by minimizing gain factor $\beta$, although this is difficult for a semiconductor laser, particularly since $\beta$ can scale inversely with the size of the gain medium  \cite{Ma2018}. Obtaining a gain chip or antireflective coated laser diode with the right parameters is challenging, especially for green wavelengths. This problem also exists for infrared absorption, since the laser emission must match the 1042 nm gap in the singlet state. In our scheme, running in the infrared could be achieved by extending the external cavity design proposed here using a diffraction grating in a Littrow and Littman–Metcalf configuration to create a tunable system. 

We consider in this work a normal incidence beam path through a fixed diamond thickness $d$=500$\mu$m. We note that a higher sensitivity within feasible limits of threshold current could potentially be reached with a thinner diamond with a very high NV density. However, in the limit of $d$$\rightarrow$0, other effects not considered in our model may act to limit performance, such as variation in NV density and the role of other types of defects. Measurements of absorption and $T_2^{*}$ as a function of diamond thickness would be extremely useful in determining behavior in this regime. We also consider that it may be possible to reach higher sensitivity in the low NV density regime using an extended beam path achieved through internal reflection in a thicker diamond. This again requires new experimental measurements to precisely quantify losses (due to reflection or absorption) in such a geometry. 

We note that the fundamental limit for the scheme is the level of contrast $C$ generated between the on/off microwave resonance states, very low for a large diamond ensemble. However, the scheme is not specifically limited to diamond and is broadly applicable for any material where a large enough, controllable difference in optical absorption could be generated. The advantage of using diamond is the ability to coherently manipulate the desired states in a quantum sensing scheme. Our calculations indicate the scheme will likely only work for diamonds with a high ($>$ 1 ppm) NV$^{-}$ density. Such diamonds have a worse ensemble $T_2^{*}$ time, limited by nitrogen spin interaction. A developing solution here may be to use optimal control methods in order to better control the ensemble. Such methods are widely implemented for nuclear magnetic resonance and electron spin resonance on bulk samples, but have yet to be fully developed for sensing using diamond defects  \cite{Nbauer2015}. 

\section{Acknowledgments}

The work presented here was funded by the Novo Nordisk foundation through the synergy grant bioQ and the bigQ Center funded by the Danish National Research Foundation (DNRF).

\bibliographystyle{apsrev}
\bibliography{threshrefsb.bib}

\end{document}